\documentstyle[emulateapj]{article}

\slugcomment{To appear in the PASP (2000 Dec issue)}

\begin{document}

\title{SN 1997bs in M66:  Another Extragalactic $\eta$ Carinae 
Analog?\footnote{Based in part on observations
made with the NASA/ESA {\sl Hubble Space Telescope}, obtained from
the data archive of the Space Telescope Science Institute, which is
operated by the Association of Universities for Research in Astronomy,
Inc., under NASA contract NAS 5-26555.}}

\author{Schuyler D.~Van Dyk} 
\affil{IPAC/Caltech, Mailcode 100-22, Pasadena CA  91125}
\authoremail{vandyk@ipac.caltech.edu}

\author{Chien Y.~Peng}
\affil{Steward Observatory, Astronomy Department, University of Arizona, Tucson, AZ  85721}
\authoremail{cyp@as.arizona.edu}

\author{Jennifer Y.~King, Alexei V.~Filippenko, Richard R.~Treffers, Weidong Li}
\affil{Astronomy Department, University of California, Berkeley, CA  94720-3411}
\authoremail{jyking@ugastro.berkeley.edu, alex@wormhole.berkeley.edu, 
treffers@astro.berkeley.edu, weidong@astro.berkeley.edu}

\and

\author{Michael W.~Richmond}
\affil{Physics Department, Rochester Institute of Technology, Rochester, NY 14623}
\authoremail{mwrsps@rit.edu}

\begin{abstract}
We report on SN 1997bs in NGC 3627 (M66), the first supernova discovered by
the Lick Observatory Supernova Search using the 0.75-m Katzman Automatic Imaging
Telescope (KAIT).  Based on its early-time optical spectrum, SN 1997bs was 
classified as Type IIn.  However, from the $BVRI$ 
light curves obtained by KAIT early in the supernova's evolution, and F555W and 
F814W light curves obtained from {\sl Hubble Space Telescope\/} archival WFPC2 
images at late times, we question the identification of SN 1997bs as a 
{\it bona fide\/} supernova.  We believe that it is more likely a 
super-outburst of a very massive luminous blue variable star, analogous to 
$\eta$ Carinae, and similar to SN 1961V in NGC 1058 (Filippenko et 
al.~1995 [AJ, 110, 2261]) and SN 1954J (``Variable 12'') in NGC 2403
(Humphreys \& Davidson 1994 [PASP, 106, 1025]).  
The progenitor may have survived the outburst, 
since the SN is seen in early 1998 at $m_{\rm F555W}=23.4$, about 0.5 mag 
fainter than the progenitor identified by Van Dyk et al.~(1999, [AJ, 118,
2331]) in a pre-discovery image.  Based on analysis of its environment in the 
{\sl Hubble Space Telescope\/} images, the progenitor was not in an H~II region
or association of massive stars.  The recent discovery of additional
objects with properties similar to those of SN 1997bs suggests that the 
heterogeneous class of Type IIn supernovae consists in part of ``impostors.''
\end{abstract}


\section{Introduction}

The ultimate fate of the most massive stars in our Galaxy and other galaxies is
still not well understood.  The post-main-sequence evolution of stars with 
masses $M \gtrsim 20$--$30\ M_{\sun}$, theoretically, should take them possibly 
through the red supergiant phase, or directly to the luminous blue variable 
(LBV) phase, en route to becoming hydrogen-deficient or hydrogenless Wolf-Rayet 
stars toward the end of their lives (e.g., Langer et al.~1994; Stothers \& Chin 
1996).  Although the number of known examples of LBVs is small (Humphreys \& 
Davidson 1994), the spectacular cases of $\eta$ Carinae (e.g., Davidson \& 
Humphreys 1997) and the Pistol Star (Figer et al.~1998, 1999) show that they go
through spectacular eruptive phases of mass ejection.  It is assumed that very 
massive stars ultimately explode as supernovae (SNe; e.g., Woosley, Langer, \& 
Weaver 1993), but it is still possible that they ``fail'' to become supernovae
and core collapse directly to form black holes (MacFadyen \& Woosley 1999).

The spectroscopic characteristics of SNe have been recently reviewed by Filippenko
(1997, 2000a).  In brief, Type I SNe (SNe I), lacking hydrogen in their optical 
spectra, divide into the classical SNe Ia, which probably arise from exploding 
white dwarf stars, and SNe Ib/c, which appear to be more closely related to 
Type II SNe.  All SNe II show hydrogen in their optical spectra and are 
thought to arise from the explosions of massive stars 
($M \gtrsim 8\ M_{\sun}$).  Great variation exists in the strength and profile 
of the hydrogen lines of SNe II.  A growing number of ``peculiar'' SNe II show 
a hydrogen emission-line combination of a narrow component atop a broader
component (sometimes having a complex shape), with no P-Cygni-like absorption 
trough.  These are the SNe II-``narrow'' (SNe IIn; Schlegel 1990; Filippenko 
1991).  

As of 2000 May, 43 SNe IIn have been identified (from the online Asiago SN
Catalog\footnote{http://merlino.pd.astro.it/$\sim$supern/snean.txt} and the IAU 
Circulars).  The Balmer emission has been 
interpreted as arising from interaction of the SN ultraviolet radiation and the
SN ejecta with a dense circumstellar medium set up by the progenitor in a 
pre-SN phase.  Evidence is accumulating that SNe IIn arise from very massive 
stars (e.g., Weiler, Panagia, \& Sramek 1990; Sollerman, Cumming, \& Lundqvist 
1998).  It seems likely that the old Zwicky (1965) ``Type V'' SNe, examples of 
which include SNe 1954J and 1961V, and which are now generally considered to 
have been peculiar SNe II (Doggett \& Branch 1985; Filippenko 1991), are 
related to SNe IIn.  For instance, SN 1986J, a prototypical SN IIn, was 
originally classified as Type V (Rupen et al.~1987).  However, particularly
in the case of SN 1961V (Goodrich et al.~1989; Filippenko et al.~1995),
it appears that not all of the Zwicky types were genuine SNe, strictly defined 
to be the violent destruction of a star at the end of its life.  By extension,
then, it is intriguingly possible that not all SNe IIn are ``real'' SNe.  Some 
SNe IIn could represent a pre-SN cataclysmic event of a very massive star.  
This ``contamination'' of the SN IIn subtype complicates calculations of the 
production rate of neutron stars and the chemical evolution of galaxies, and
adds a new wrinkle in the evolution of the most massive stars.

In this paper, we discuss the example of the SN IIn 1997bs in NGC 3627 (M66).
SN 1997bs was the first SN discovered by the Lick Observatory Supernova Search 
(LOSS\footnote{http://astro.berkeley.edu/$\sim$bait/kait.html}, the direct 
successor of the former search at Leuschner Observatory), 
which is robotically conducted with the 0.75-m Katzman Automatic Imaging 
Telescope (KAIT; see Filippenko et al.~2000).  Treffers et al.~(1997) reported 
the SN discovery, from an image (Figure 1) obtained on 1997 April 15 (UT
dates are used throughout this paper).  The SN brightness at discovery was 
roughly estimated to be $R \simeq 17$ mag.  An image from April 10 did not 
show anything at the location of SN 1997bs, to a limiting magnitude $R \simeq 17.6$.
The position of the SN is RA=$11^h 20^m 14{\farcs}25$, 
Dec=$+12\arcdeg 58\arcmin 19{\farcs}6$ (J2000.0; Cavagna 1997), which is about 
11${\farcs}$2 west and 69${\farcs}$9 south of the galaxy's nucleus.  
NGC 3627 has been host to two other known SNe, the Type II SN 1973R (Ciatti \& 
Rosino 1977) and the Type Ia SN 1989B (Wells et al.~1994).  Here we present 
evidence that SN 1997bs was not a genuine SN, through spectroscopic and
photometric observations made at Lick and also photometry from archival
{\sl Hubble Space Telescope\/} ({\sl HST}) images.

\section{Observations}

\subsection{Lick Spectroscopy}

On 1997 April 16, CCD spectra of SN 1997bs were obtained by Filippenko, Barth,
\& Gilbert (1997) with the Kast spectrograph (Miller \& Stone 1993) on the
Shane 3-m reflector at Lick Observatory.  The slit width was 2\arcsec, and
three different gratings and grisms were used to span the wavelength range
3200--8000 \AA\ with a typical resolution of about 7 \AA.  The spectra were
extracted and calibrated following standard procedures, and we took special care
to remove telluric absorption features through division by the spectrum of the
flux comparison star.  Owing to variable seeing, the absolute flux calibration
is only approximate, but the shape of the spectrum should be reasonably 
accurate.  

The spectrum (Figure 2) is dominated by
relatively narrow (FWHM $\simeq$ 1000 km s$^{-1}$) Balmer emission lines on a 
featureless continuum.  Many weaker Fe II emission lines are also evident.
From this spectrum, Filippenko et al.~(1997) classified SN 1997bs as Type IIn.  

\subsection{KAIT Photometry}

$BVRI$ photometry was obtained with KAIT over about 60 days following 
discovery before it was no longer detectable.  
Images of NGC 3627 in each of the four bands were obtained on 1994 
January 13 at the KPNO 2.1-m telescope by L.~A.~Wells (who kindly provided 
them to us).  These pre-SN images served as ``templates'' which were subtracted 
from the SN images in each of the four bands 
(see Filippenko et al.~1986; Richmond et al.~1995).  First, for
each band, the template was registered to a SN image, so that stars in the 
field were aligned, and the template image was convolved with a Gaussian 
to match the FWHM of stars in the SN image.  Then, the stars in the template 
image were flux-scaled to match the stars in the SN image, and, finally, the 
template was subtracted from the SN image, leaving behind a residual image 
containing only the SN.  This procedure was followed for each observational 
epoch.  Photometry of the SN in each band and for each epoch was done on the 
template-subtracted image using the APPHOT package in 
IRAF\footnote{IRAF (Image Reduction and Analysis Facility) is distributed by the 
National Optical Astronomy Observatories, which are operated by the Association of 
Universities for Research in Astronomy, Inc., under cooperative agreement with the 
National Science Foundation.}.  Measurements of the field stars ``3'' and 
``4'' (see Wells et al.~1994) were also obtained from the original (not 
subtracted) images in each band for each epoch, this time determining the sky 
background from an annulus around each field star.

In addition, $BVRI$ images of NGC 3627 were obtained on 1999 December 17 
with the 1.5-m telescope at Palomar Observatory, under photometric conditions.
In Table 1 we list the magnitudes in each band for stars ``3'' and ``4''
derived from the Palomar observations.
The agreement with the values given in Table 2 of Wells et al.~(1994) for
these stars is quite good.  With our magnitudes for these stars we then
transformed the KAIT photometry of SN 1997bs onto the Johnson-Cousins 
system.

\subsection{{\sl HST\/} Archival Image Photometry}

Deep {\sl HST\/} WFPC2 images of NGC 3627 in the bands F555W and F814W were 
obtained from the image archive.  These images were made in pairs (to 
facilitate removal of cosmic rays) between late 1997 and early 1998 during 
several epochs.  They were originally intended to measure the distance to the 
galaxy using Cepheid variables (GO program 6549), with the goal of determining
the absolute magnitude of SN 1989B (Saha et al.~1999; their Table 1 lists the 
exposures).  Saha et al.~identify SN 1997bs as variable star ``C2-V23,'' yet
they did not recognize it as the SN and note only that
C2-V23 and others ``are definitely variable ...'' and ``...may be 
transient variables such as novae.''

A cosmic-ray split pair of WFPC2 F606W images made on 1994 December 28
was also obtained from the archive.  Van Dyk et al.~(1999; their Figure 7) 
display this image pair, as well as representative F555W images showing the SN.

To measure the brightness of SN 1997bs in the WFPC2 images, 
we employed point-spread function (PSF) fitting photometry performed by 
DAOPHOT (Stetson 1987) and ALLSTAR within IRAF.
We used the {\sl Tiny Tim\/} routine (Krist 1995) to produce an artificial 
point-spread function (PSF) image, simulated to have
stellar spectral type A, image radius $0{\farcs}5$, and full width at half
maximum of $\sim 0{\farcs}07$.
The artificial PSF was created at the pixel position of the SN in the archive
images.  Since the stellar profiles in the images were actually somewhat
broader than the artificial PSF, the latter was broadened using a Gaussian
to match the former.  Instrumental photometry was aperture corrected, using
measurements of artificial stars of known magnitude placed randomly in the 
field, and converted to magnitudes using the 
photometric synthetic zeropoints in Holtzman et al.~(1995; their Table 9), 
appropriate for a $0{\farcs}5$ aperture radius.  A correction 
for the gain ratio to the 7 $e^-$ ADU$^{-1}$ gain state was applied.

\section{Results}

In Table 2 we list the KAIT photometry for SN 1997bs, while
Table 3 gives the photometry derived from the archival {\sl HST\/}
F555W and F814W images.  Our {\sl HST\/} photometry of SN 1997bs agrees quite
well with that from Saha et al.~(1999; their Table 3), to within 0.02 mag at
F555W and 0.04 mag at F814W.

In Figure 3 we show the combined $BVRI$ light curves for the SN, including
both the KAIT and {\sl HST\/} photometry.
The F555W and F814W measurements correspond approximately to $V$ and $I$
magnitudes, respectively.  These WFPC2 synthetic magnitudes were {\it not\/}
converted to the Johnson-Cousins system following (for example) the 
transformations given by Holtzman et al.~(1995); for an emission-line 
object such as SN 1997bs, this conversion may not be entirely correct or 
appropriate.

SN 1997bs was apparently discovered before maximum brightness in $R$, and 
generally at or near maximum in the other bands.  The SN light curves decline 
from peak brightness over the next $\sim$20 days by $\sim$1 mag.  The decline  
in $V$ is
less drastic after about JD 2450580.  In $R$ and $I$, the light curves
appear to reach a plateau of roughly constant magnitude in each band.  (The
observational coverage in $B$ terminates after JD 2450582.)  The KAIT photometry
in all bands ends on JD 2450617, after a span of nearly 60 days of monitoring.

There is a long, $\sim$150-day gap in our coverage before we are able to extend
the light curves using the {\sl HST\/} photometry in (approximately) $V$ and
$I$ bands.  It is apparent in $V$ that the light curve began a much steeper
decline on or before JD 2450765; the slope of the decline is quite similar
to the slope of the earlier post-maximum decline.  The $V$ light curve, again,
achieves a more gradual decline after about JD 2450786.  In fact, after JD 
2450804, the $V$ magnitude of the SN appears to reach another plateau, at 
$\sim$23.4 mag.
The $I$ light curve at late times appears to gradually decline, at a slower
rate than the earlier post-maximum decline (the SN is
partially saturated in the first-epoch F814W image, so we can only give a 
lower limit to its brightness).

In Figure 4 we show the color evolution of the SN.  For the first 60 days,
the SN color becomes generally redder after discovery.  This is best 
exemplified by the evolution of the $V-I$ color, which near maximum already
is at $V-I \simeq 0.62$ mag, but, even before the end of KAIT monitoring, has
reached $V-I \simeq 1.43$ mag.  By the time monitoring resumes, with the 
inclusion of the {\sl HST\/} data, the SN is very red, at $V-I \approx 3.4$ mag.
The red color of the SN is obvious in the color composite image shown by Saha 
et al.~(1999; their Figure 1).  The SN possibly may have been much redder than 
this prior to its recovery in the {\sl HST\/} images:  it becomes bluer at 
late times, to $V-I \approx 3$ mag in the last F555W/F814W image pair.

\section{Discussion}

Although the spectrum of SN 1997bs indicates that it is a SN IIn, its 
photometric nature is unusual, both in the behavior of the light curves
and in absolute magnitude.  

Since the distance modulus of 
NGC 3627 has been measured with Cepheids ($\mu=30.28$ mag; Saha et 
al.~1999), we can adjust our observed light curves of SN 1997bs to an absolute
scale.  We would like to correct the absolute light curves for extinction, but
the amount of reddening toward SN 1997bs is not known.  The Galactic foreground 
reddening toward NGC 3627 is $E(B-V)=0.03$ mag (Schlegel et al.~1998).  
However, SN 1997bs occurred at the edge of a dust lane in NGC 3627, so it is 
likely that reddening internal to the galaxy at the site of the SN is
appreciable and spatially variable.  We have attempted to estimate the 
reddening, based on the colors of stars in the SN environment (see below), yet 
the stellar colors, particularly for the bluest stars, are consistent with the 
foreground reddening estimate.  

From the light curves, we find that the $B-V$ color near maximum was 0.67 mag. 
If the intrinsic color was close to $B-V=0.0$, which is what one might expect 
for SNe II, then $E(B-V)\simeq 0.7$ mag.  However, the spectrum of SN 1997bs 
dereddened by this amount would make the SN continuum unphysically blue, so 
already we know
the reddening cannot be this large.  The emission-line intensity ratios of 
H$\alpha$/H$\beta$/H$\gamma$ are 2.6/1.0/0.5, which, if anything, are slightly
flatter than what one expects for Case B recombination (2.8/1.0/0.47); so, 
there is no evidence for reddening of the emission lines.  We measure an 
equivalent width of 1.8 \AA\ for the narrow Na I D absorption line at the 
redshift of the SN (unfortunately, the two line components are not resolved).  
Though with low-resolution spectra one cannot confidently translate this into a
reddening (one actually needs high-dispersion spectra), an analysis
similar to that of Filippenko, Porter, \& Sargent (1990) suggests a reddening of
$E(B-V)=0.38$ mag.  Moreover, when this type of analysis was applied to the SN 
Ic 1987M by Jeffery et al.~(1991), they argued that it
overestimates $E(B-V)$ by a factor 1.8.  If we scale our estimate for SN 1997bs 
by this factor, then $E(B-V)=0.21$ mag.  We assume this value
to be the total (but quite uncertain) reddening to the SN and correct the 
observed light curves for it, using a standard Galactic reddening law 
(Cardelli, Clayton, \& Mathis 1989), $A_V=3.1 E(B-V)$.

SN 1997bs appears to be unusual with respect to most other SNe II (or other SN
types), due to its very faint absolute magnitude: maximum for the SN was at 
only $M_V \simeq -13.8$ mag, about 4 to 4.5 mag fainter than a typical SN
II.  

In Figure 5 we show the object's unusual photometric evolution through 
a comparison of the absolute $V$ light curves of SN 1997bs 
and of the SNe IIn 1988Z (Turatto et 
al.~1993), 1987B (Schlegel et al.~1996), and 1994Y (Ho et al.~2000).
(Distances to the latter three SNe were calculated from their recession
velocities, obtained from the NASA/IPAC Extragalactic Database and an
assumed distance scale $H_0=65$ km s$^{-1}$ Mpc$^{-1}$.  No extinction
corrections have been applied for these SNe.)  
The light curves of the three comparison SNe IIn have been scaled, so that 
their maximum observed brightness occurs at approximately the same epoch 
as SN 1997bs.  One can see that, besides the obvious underluminosity,
the post-maximum decline of SN 1997bs is very different in behavior from 
that of the other SNe.  

One also notices from Figure 5 that, at late times, the F555W light 
curve (in particular) appears to reach a plateau once again, at $\sim -7.5$ 
mag (or $\sim$23.4 mag from Figure 3).
None of the other SNe IIn, particularly SN 1994Y, show this sort of behavior.
Van Dyk et al.~(1999) found in the F606W image from 1994 December 28, about 
27.5 months {\it before\/} the discovery of SN 1997bs, a star at the exact 
position of the SN, with $m_{\rm F606W}=22.86 \pm 0.16$ mag.  Assuming
$m_{\rm F606W} \approx V$, the star had an absolute magnitude 
$M_V \simeq -8.1$ mag, at the distance of NGC 3627 and corrected for our assumed 
reddening.  Van Dyk et al.~associated this star with the progenitor of SN 1997bs.  
The late-time brightness of SN 1997bs is $\sim$0.5 mag fainter than
the brightness of this precursor star (see Figure 6).  
However, the unusual apparent 
flattening-out of the late-time light curve suggests that the star {\it may\/}
have survived the explosion.

This raises the question of whether SN 1997bs was really a
{\it bona fide\/} supernova, or instead was something altogether
different.  We believe that the evidence presented in this paper 
indicates that SN 1997bs was more likely the super-outburst of a very massive 
LBV star, possibly analogous to the enormous eruptions experienced by $\eta$ 
Carinae (Davidson \& Humphreys 1997).  The $V$ light curve, in particular, 
shows that SN 1997bs brightened by nearly 6 mag before fading, possibly 
through the formation of optically thick dust, given how red it became.

SN 1997bs is not alone in its unusual characteristics.
A similar object may be the peculiar SN IIn 1961V in NGC 1058.  SN 1961V was
originally classified as ``Type V'' by Zwicky (1964, 1965) and had
probably the most bizarre light curve ever recorded for a SN. 
Its progenitor was identified as a
very luminous star, $M^0_{\rm pg} \approx -12$ mag, visible in many 
photographs of NGC 1058 prior to the explosion (Bertola 1964; Zwicky 1964;
Klemola 1986).  Goodrich et al.~(1989) and Filippenko et al.~(1995) identified
the possible survivor of the SN 1961V event, the latter study via {\sl HST\/}
WFPC imaging, and both studies conclude that perhaps SN 1961V was not a 
genuine supernova, but rather also the super-outburst of a luminous blue 
variable, similar to $\eta$ Car.

We can compare the photometric behavior of SN 1997bs with
that of the $\eta$ Car-like variables discussed thoroughly by Humphreys, 
Davidson, \& Smith (1999).  In Figure 6 we show the light curve of SN 
1997bs, along with the schematic light curves for $\eta$ Car, SN 1961V, 
and SN 1954J (also known as 
``Variable 12,'' V12, in NGC 2403; Humphreys \& Davidson 1994).
(Interestingly, both $\eta$ Car and SN 1954J were originally classified as 
Type V SNe by Zwicky 1964 and, e.g., Schaefer 1996, respectively.)
$V$-band light curves are shown for SN 1997bs, $\eta$ Car 
(Humphreys et al.~1999), and SN 1954J (Tammann \& Sandage 1968), while
the light curve for SN 1961V is photographic (Humphreys et al.~1999).  
We have adjusted all of the curves to absolute magnitudes: the distances 
to SN 1961V (8 Mpc), SN 1954J (3.2 Mpc), and $\eta$ Car 
(2300 pc) are from Goodrich et al.~(1989), Freedman \& Madore (1988), 
and Davidson \& Humphreys (1997), respectively.   (The curve for SN 1997bs 
is that from Figure 5, with the addition of the 1994 pre-discovery observation; 
the curves for the other three objects 
have not been extinction-corrected.)  All of the curves have been further
adjusted, so that the maxima occur at the same time.

Overall, there are a number of similarities between these various light curves,
although there are also some notable differences.
The eruption of $\eta$ Car lasted 20 years, while the outburst of SN 1997bs, 
with its shorter duration, is much more similar to those of SN 1961V and 
possibly V12 in NGC 2403.  However, the maximum luminosity of the SN 1997bs
outburst is very much like that of the great eruption of $\eta$ Car. 
Given the short duration of its eruption, the total energy emitted
by SN 1997bs is much less than either $\eta$ Car or SN 1961V and is probably 
more like that of V12 or P Cygni (not shown here, but also discussed
by Humphreys et al.~1999).  
SN 1961V and V12 show a 
flattening (plateau) one to two years past maximum, similar to the
behavior shown by SN 1997bs, and years later faded considerably, 
probably due to the formation of dust 
in the ejected material (Goodrich et al.~1989; Filippenko et al.~1995).

From a color standpoint, LBVs, including $\eta$ Car,
have the spectra of F-type supergiants during eruption (Humphreys et al.~1999),
with intrinsic $B-V \simeq 0.3$ mag.  If this were the intrinsic color of
SN 1997bs at outburst, the reddening for SN 1997bs would 
be $E(B-V) \simeq 0.37$ mag, which is quite 
close to our estimate of the reddening based on the Na~I D 
absorption line in the optical spectrum.

What was the star that gave rise to SN 1997bs?  With $M_V \simeq -8.1$ mag,
it is too luminous to have been an O-type star and
was most likely a supergiant later in spectral type than O9 or B0.  
If the 
pre-outburst star was in an LBV quiescent state, then it was most likely a
mid B-type star, with a surface temperature of 20,000-25,000 K
and $M_{\rm bol} \approx -10$ to $-10.5$ mag, which is 
typical for these objects, with initial mass 
$\sim$60 $M_{\odot}$ and a main sequence lifetime of $\sim$4 Myr  
(Humphreys \& Davidson 1994; Humphreys et al.~1999; Stothers \& Chin 1996).

We can attempt to estimate the mass of the progenitor star from 
an environmental analysis of the properties
of other stars in its vicinity, much the same way as in Van Dyk et al.~(1999).
We have more thoroughly analyzed the {\sl HST\/} data containing SN 1997bs
by producing very deep images in both the F555W and F814W bands through 
coaddition of the individual images taken at the various epochs.   
The results are a 24500-s total exposure in F555W and
a 12500-s exposure in F814W.  We show the F555W total image in Figure 7.  
For the distance modulus $\mu=30.28$ mag measured by Saha et al.~(1999), 
1\arcsec\ = 55 pc.  For the WFPC2 chip, then, one pixel $\simeq$ 6 pc.
We note that at this resolution, it is difficult to distinguish a single 
object from a 
blend of several contiguous ones, so it is possible that the SN progenitor
could have exploded among a small, compact star cluster, and that now we see
the surviving cluster minus the progenitor.  With the available data, we 
cannot rule out this possibility.

We have analyzed the environment of SN 1997bs by measuring the magnitudes, 
using the PSF-fitting techniques described in \S 2.3, of the stars in a 
20\arcsec\ (north-south) $\times$ 12\arcsec\ (east-west) region
centered on SN 1997bs.  Figure 8 shows the resulting 
color-magnitude diagram.  Overlaid are the theoretical isochrones for 
solar metallicity from Bertelli et al.~(1994), after conversion, as in
Van Dyk et al.~(1999), from the Johnson-Cousins system to the WFPC2 synthetic 
magnitude system.  We also converted the reddening and extinction to the WFPC2
synthetic system.  

Although many young (massive) dwarf and supergiant stars are 
resolved in this larger area, and the abundance of dust and luminous stars in 
the environment implies recent massive star formation,
the stars within $\sim$1\arcsec\ ($\sim$55 pc) of SN 1997bs are older, less 
luminous giant and supergiant stars.  Regardless of whether
the progenitor of SN 1997bs has survived the outburst, the star
(or possible surviving compact star cluster) does not seem to have 
formed in an H II region or an association of obvious O-type stars,
unlike the case of $\eta$ Car, which formed
among clusters of young massive stars of 
similar age (Davidson \& Humphreys 1997).  However, P Cygni, which may be
similar to the progenitor of SN 1997bs, is in the Cyg OB1 association, a group
of mostly B-type stars.

Based on the diagram in Figure 8, we are unable to place a meaningful 
constraint on the mass of the SN 1997bs progenitor.  To perform a more 
complete assessment of the massive-star population in the vicinity of SN 1997bs,
and to determine whether the progenitor alone survived, or is among a 
surviving star cluster, higher-resolution $U$ and $B$ images must be obtained,
since, in particular, $V-I$ is a poor discriminant of hot stars.
These blue images would also provide a better estimate of the extinction 
toward the SN.

Two more recent examples of objects similar to SN 1997bs
possibly include SN 1999bw in NGC 3198 (Filippenko, Li, \& 
Modjaz 1999) and 
SN 2000ch in NGC 3432 (Filippenko 2000b).  SN 1999bw appears to have similar 
spectral characteristics to SN 1997bs and is also quite subluminous.  SN
2000ch is similar, but may have experienced a shorter-lived outburst
than did SN 1997bs (Hudec et al.~2000).

The SN IIn subclass clearly spans a very broad range of properties, as was
already known to some extent (e.g., Filippenko 1997).  If the conclusions
of this paper are correct, and the progenitor of SN 1997bs survived, then
the ``explosion'' mechanism is {\it not\/} core collapse in all SNe IIn.
That is, the subclass appears to include a number of supernova ``impostors'' 
such as SN 1997bs, which are evolved but still living.  For SN 1997bs, a more 
definitive test will be whether the star remains visible in future
{\it HST\/} images obtained years after the outburst (although, even if it
survived, it might fade at optical wavelengths, due to the formation of dust
in the ejecta; see Goodrich et al.~1989).  The occurrence of 
SN 1997bs and its possible cousins (SNe 1961V, 1954J, 1999bw, and 2000ch), and 
their resemblance to $\eta$ Car and LBV super-outbursts, begs the question of 
the ultimate fate of these very massive stars.

\acknowledgements

The work of A.V.F.'s group at UC Berkeley is supported by NSF grants
AST-9417213 and AST-9987438, as well as by NASA grant AR-08006 from the Space 
Telescope Science
Institute, which is operated by AURA, Inc., under NASA contract NAS5-26555.
KAIT was made possible by generous donations from Sun Microsystems, Inc.,
Photometrics, Ltd., the Hewlett-Packard Company, AutoScope Corporation,
the National Science Foundation, Lick Observatory, the University of 
California, and the Sylvia and Jim Katzman Foundation.  We thank A.~J.~Barth, 
D.~C.~Leonard,
and A.~M.~Gilbert for assistance with the Lick 3-m observations and reductions.
We are also grateful to L.~A.~Wells for providing template CCD images of NGC 
3627.


\clearpage

\begin{deluxetable}{lcccc}
\def\phmm{\phm{$-$}}
\tablenum{1}
\tablecolumns{5}
\tablecaption{Comparison Star Magnitudes and Colors}
\tablehead{\colhead{Star\tablenotemark{a}} & \colhead{$V$} & \colhead{$B-V$}
& \colhead{$V-R$} & \colhead{$R-I$}}
\startdata
3 &   16.24 &  0.53 &  0.32  &  0.37 \nl
4 &   15.95 &  0.61 &  0.36  &  0.41 \nl
\enddata
\tablenotetext{a}{The star numbers are those designated by Wells et 
al.~(1994).}
\end{deluxetable}

\clearpage

\begin{deluxetable}{llcccc}
\def\phmm{\phm{$-$}}
\tablenum{2}
\tablecolumns{5}
\tablecaption{KAIT Photometry of SN 1997bs\tablenotemark{}}
\tablehead{\colhead{UT Date} & \colhead{JD} & \colhead{$B$} & \colhead{$V$} & 
\colhead{$R$} & \colhead{$I$}}
\startdata
1997 Apr 10.27 & 2450548.77 & \nodata    & \nodata    & \llap{$>$}17.6 & \nodata \nl
1997 Apr 15.21 & 2450553.71 & \nodata    & \nodata    & 16.93(03) & \nodata    \nl
1997 Apr 16.27 & 2450554.77 & 17.79(11) & 17.12(04) & 16.85(02) & 16.45(03) \nl
1997 Apr 22.31 & 2450560.81 & 18.06(12) & 17.22(07) & 16.83(03) & 16.48(04) \nl
1997 Apr 26.22 & 2450564.72 & 17.88(05) & 17.18(03) & 16.87(02) & 16.46(02) \nl
1997 Apr 28.22 & 2450566.72 & \nodata    & 17.28(04) & 16.97(02) & \nodata    \nl
1997 Apr 30.21 & 2450568.71 & 18.04(07) & 17.43(06) & 17.01(03) & \nodata    \nl
1997 May 02.30 & 2450570.80 & 18.07(10) & 17.40(04) & 17.06(02) & 16.57(05) \nl
1997 May 03.24 & 2450571.74 & \nodata    & 17.57(17) & 17.21(08) & 16.73(05) \nl
1997 May 04.20 & 2450572.70 & 18.20(18) & 17.61(07) & 17.16(04) & \nodata    \nl
1997 May 05.20 & 2450573.70 & 18.38(16) & 17.53(11) & 17.30(03) & \nodata    \nl
1997 May 06.26 & 2450574.76 & \nodata      & \nodata  & 17.46(04) & \nodata    \nl
1997 May 07.20 & 2450575.70 & 18.80(08) & 17.85(05) & 17.59(02) & 17.16(04) \nl
1997 May 10.33 & 2450578.83 & 18.97(20) & 18.12(06) & 17.81(03) & 17.28(04) \nl
1997 May 11.31 & 2450579.81 & 19.01(06) & 18.25(05) & 17.89(02) & 17.37(04) \nl
1997 May 12.20 & 2450580.70 & 19.22(15) & 18.31(06) & 17.93(03) & 17.46(04) \nl
1997 May 13.21 & 2450581.71 & 19.18(21) & 18.24(08) & 17.89(04) & 17.39(04) \nl
1997 May 20.21 & 2450588.71 & \nodata    & 18.44(12) & 17.84(04) & 17.21(05) \nl
1997 May 21.26 & 2450589.76 & \nodata    & 18.52(14) & \nodata    & 17.41(06) \nl
1997 May 31.23 & 2450599.73 & \nodata    & 18.56(08) & 17.86(03) & 17.27(04) \nl
1997 Jun 06.24 & 2450605.74 & \nodata    & \nodata    & 17.89(03) & \nodata    \nl
1997 Jun 07.26 & 2450606.76 & \nodata    & 18.68(11) & \nodata    & 17.25(03) \nl
1997 Jun 11.21 & 2450610.71 & \nodata    & \nodata    & \nodata    & 17.32(03) \nl
1997 Jun 15.22 & 2450614.72 & \nodata    & \nodata    & \nodata    & 17.10(12) \nl
1997 Jun 17.22 & 2450616.72 & \nodata    & 18.75(12) & 18.07(05) & \nodata    \nl
\enddata
\tablenotetext{}{Note: The values given in parentheses are the uncertainties in the last two digits of the magnitudes.}
\end{deluxetable}

\clearpage

\begin{deluxetable}{llcc}
\def\phmm{\phm{$-$}}
\tablenum{3}
\tablecolumns{3}
\tablecaption{{\sl HST\/} Photometry of SN 1997bs\tablenotemark{}}
\tablehead{\colhead{UT Date} & \colhead{JD} & \colhead{$m_{\rm F555W}$} 
& \colhead{$m_{\rm F814W}$}}
\startdata
1997 Nov 12.67 & 2450765.14 & 21.45(03) & \llap{$<$}18.50 \nl
1997 Nov 16.47 & 2450768.97 & 21.73(03) & \nodata         \nl
1997 Nov 20.04 & 2450773.54 & 22.10(03) & \nodata         \nl
1997 Nov 28.10 & 2450780.60 & 22.50(03) & \nodata         \nl
1997 Dec 03.95 & 2450786.45 & 22.89(04) & 19.49(02)       \nl
1997 Dec 08.78 & 2450791.28 & 22.96(05) & \nodata         \nl
1997 Dec 14.16 & 2450796.66 & 23.05(05) & \nodata         \nl
1997 Dec 18.80 & 2450801.30 & 23.17(05) & 19.94(02)       \nl
1997 Dec 21.89 & 2450804.39 & 23.30(06) & \nodata         \nl
1997 Dec 27.67 & 2450810.17 & 23.34(05) & 20.09(02)       \nl
1998 Jan 03.53 & 2450817.03 & 23.38(06) & \nodata         \nl
1998 Jan 10.46 & 2450823.96 & 23.40(05) & 20.35(05)       \nl
\enddata
\tablenotetext{}{Note: The values given in parentheses are the uncertainties in the last two digits of the magnitudes.}
\end{deluxetable}

\clearpage

\begin{figure}
\figurenum{1}
\caption{$R$-band discovery image of SN 1997bs taken on 1997 Apr 15 with the
Katzman Automatic Imaging Telescope (KAIT), a 0.75-m fully robotic reflector
at Lick Observatory.  SN 1997bs was the first supernova discovered by KAIT.
The arrow points to the SN.  Stars 3 and 4 from Wells et al.~(1994) and Table 
1 are also indicated.}
\end{figure}


\begin{figure}
\figurenum{2}
\plotone{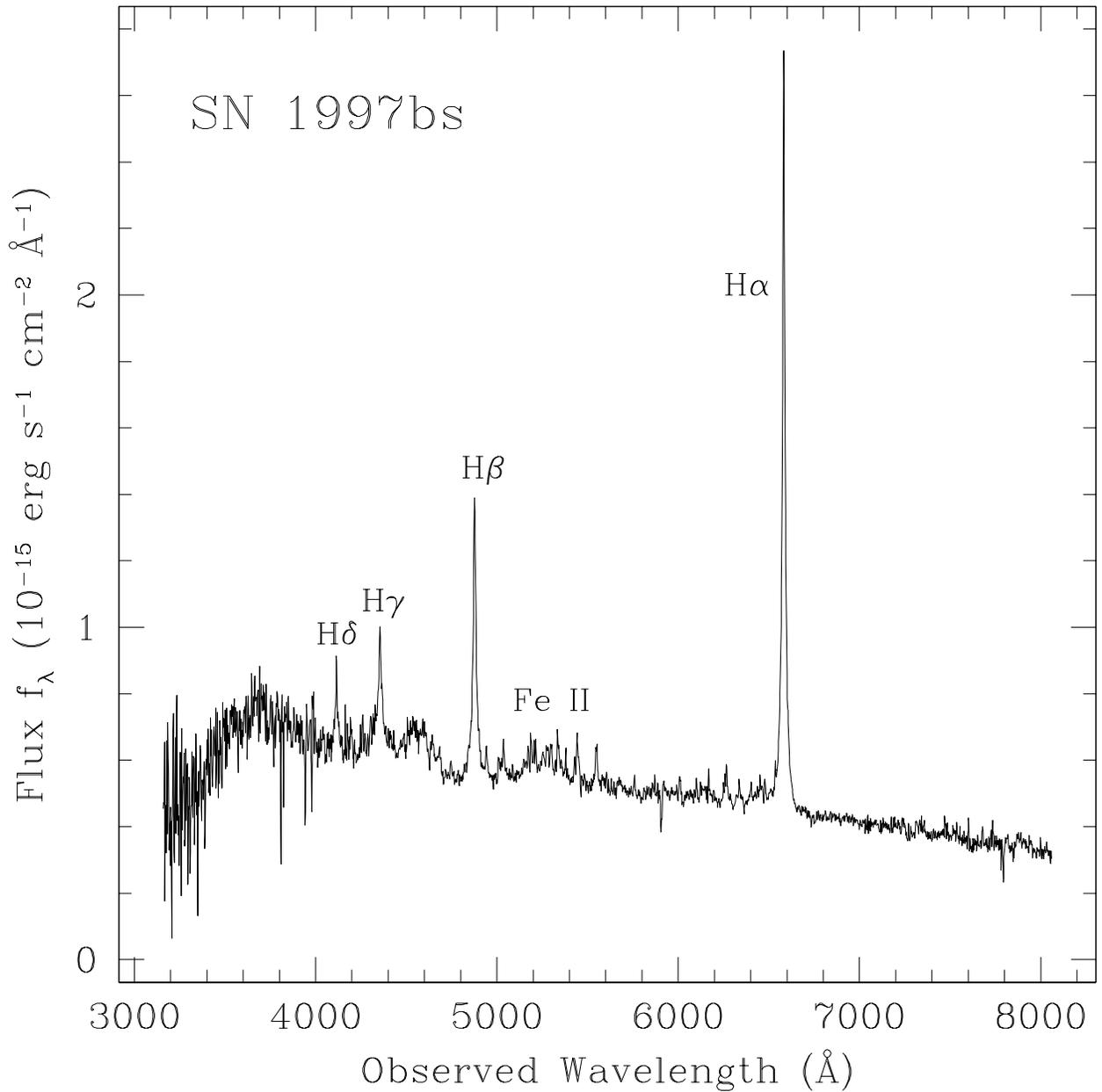}
\caption{Spectrum of SN 1997bs obtained with the Lick 3.0-m Shane reflector
on 1997 April 16.  The absolute flux scale is approximate.}
\end{figure}

\clearpage

\begin{figure}
\figurenum{3}
\plotone{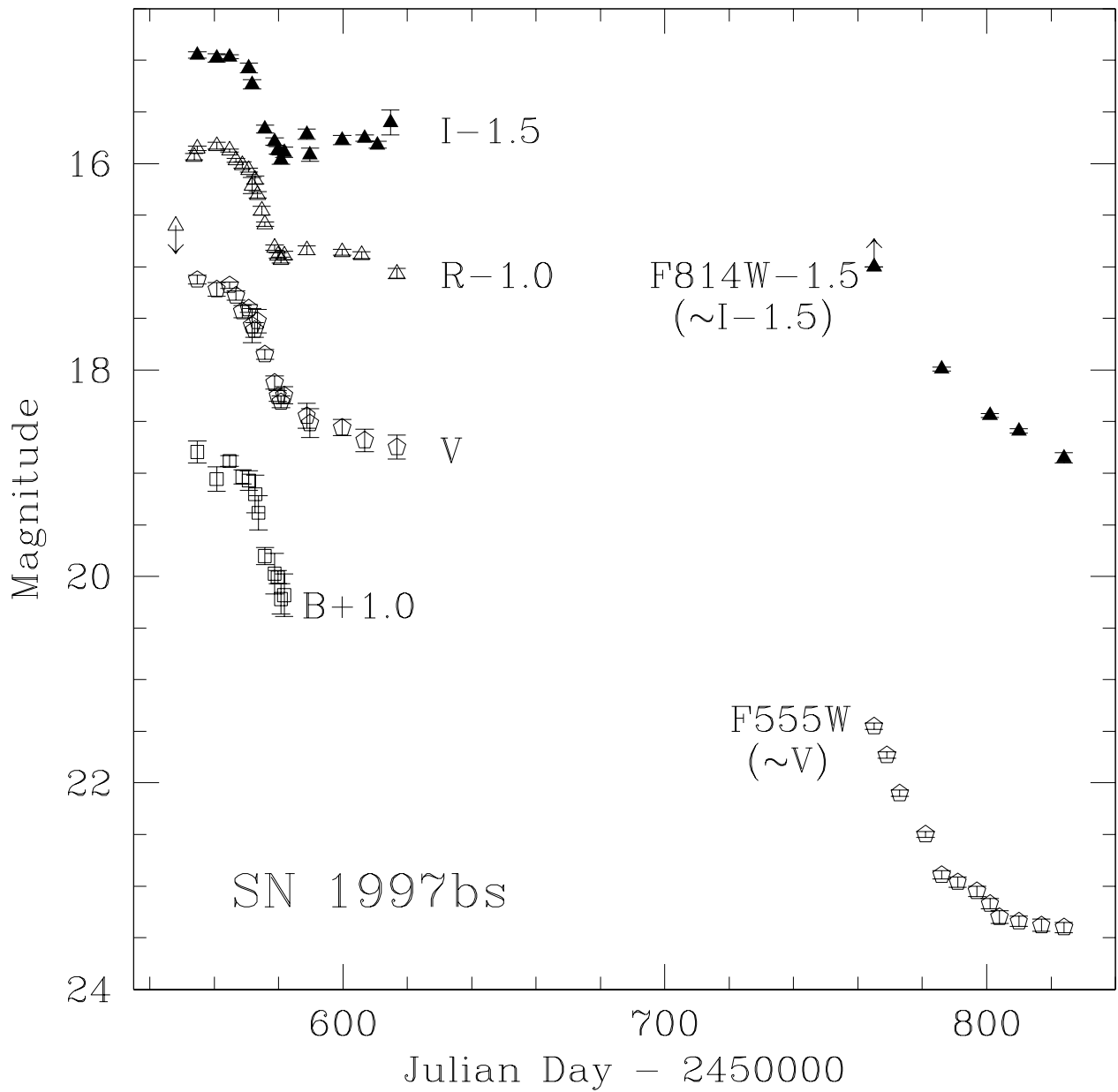}
\caption{$BVRI$ light curves for SN 1997bs, obtained with KAIT at Lick
Observatory, with the
addition of the F555W (${\sim}V$) and F814W (${\sim}I$) magnitudes obtained 
from archival {\sl HST\/} images.}
\end{figure}

\clearpage

\begin{figure}
\figurenum{4}
\plotone{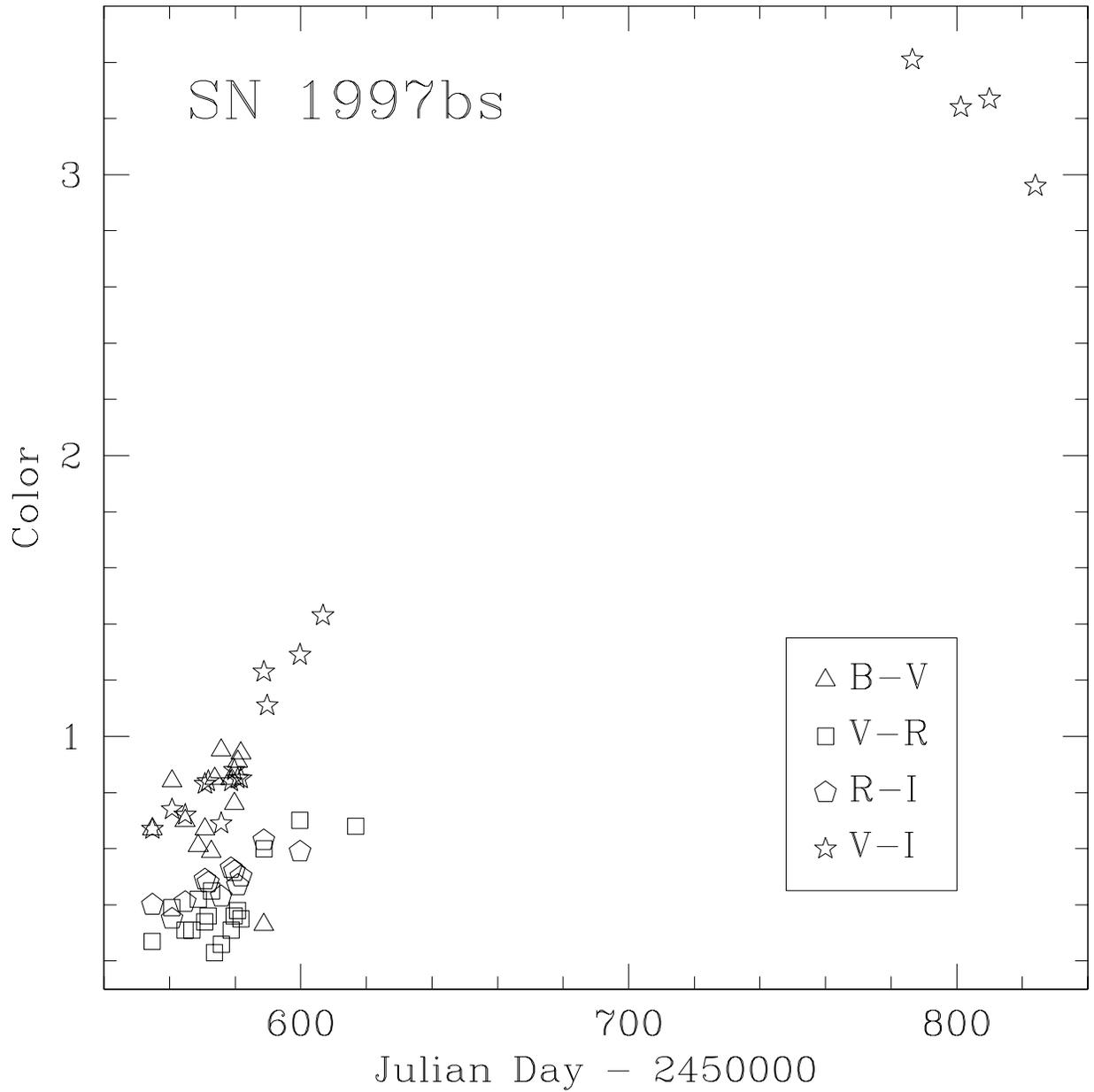}
\caption{The color evolution of SN 1997bs.  Shown are the $B-V$ 
({\it triangles}), $V-R$ ({\it squares}), $R-I$ 
({\it pentagons}), and $V-I$ ({\it stars}) colors.}
\end{figure}

\clearpage

\begin{figure}
\figurenum{5}
\plotone{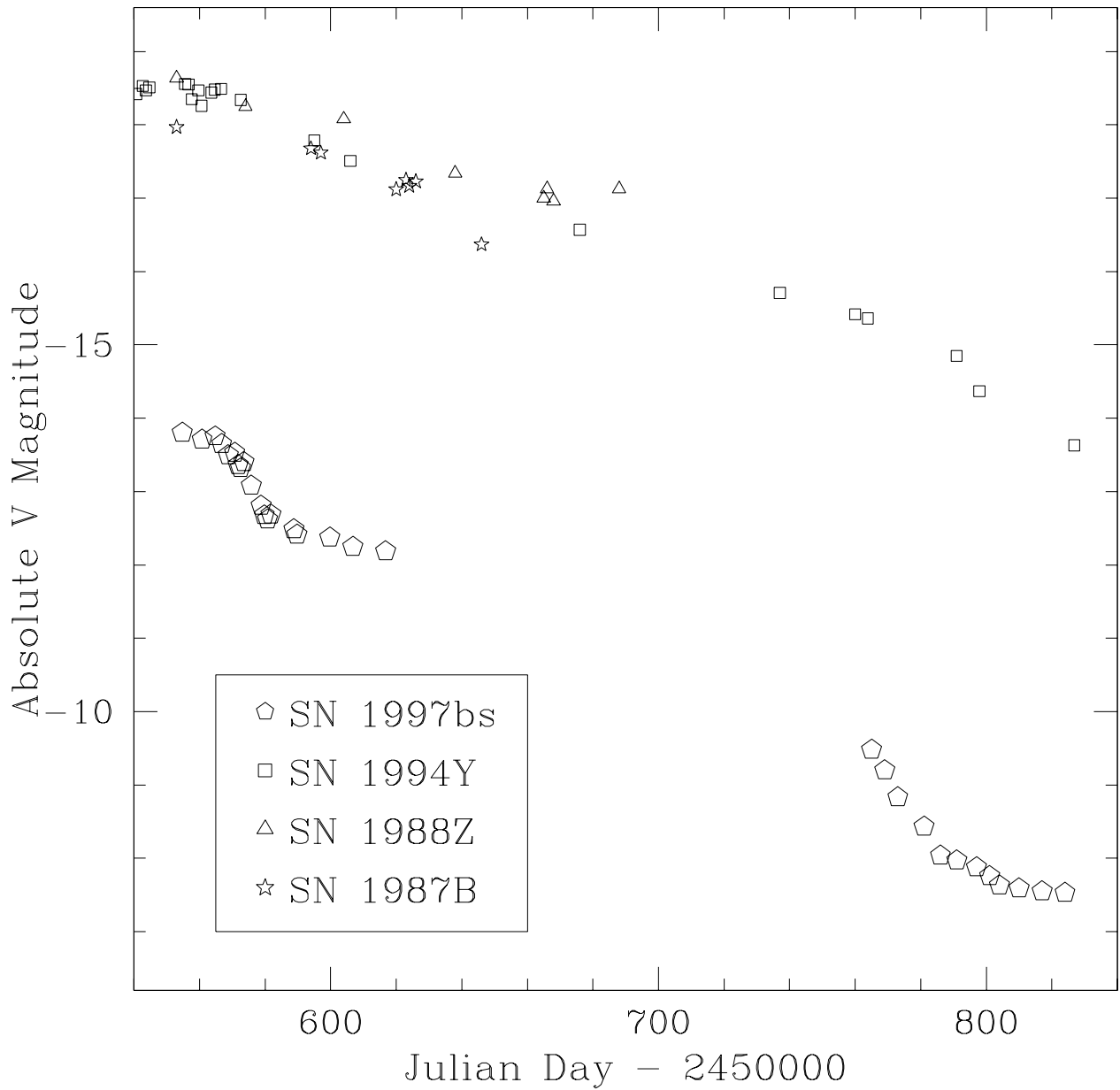}
\caption{A comparison of the absolute $V$ light curves of SN 1997bs (based on 
the KAIT and {\sl HST\/} Cepheid observations; {\it pentagons}) with those of 
three other Type IIn SNe:  SN 1994Y ({\it squares}; Ho et al.~2000); SN 1988Z 
({\it triangles}; Turatto et al.~1994); and SN 1987B ({\it stars}; Schlegel
et al.~1996).  For SN 1997bs, we assume a distance modulus $\mu=30.28$ mag
(Saha et al.~1999) and have dereddened the light curve by $E(B-V)=0.21$ mag.  
Note the relatively low luminosity of SN 1997bs.}
\end{figure}

\clearpage

\begin{figure}
\figurenum{6}
\plotone{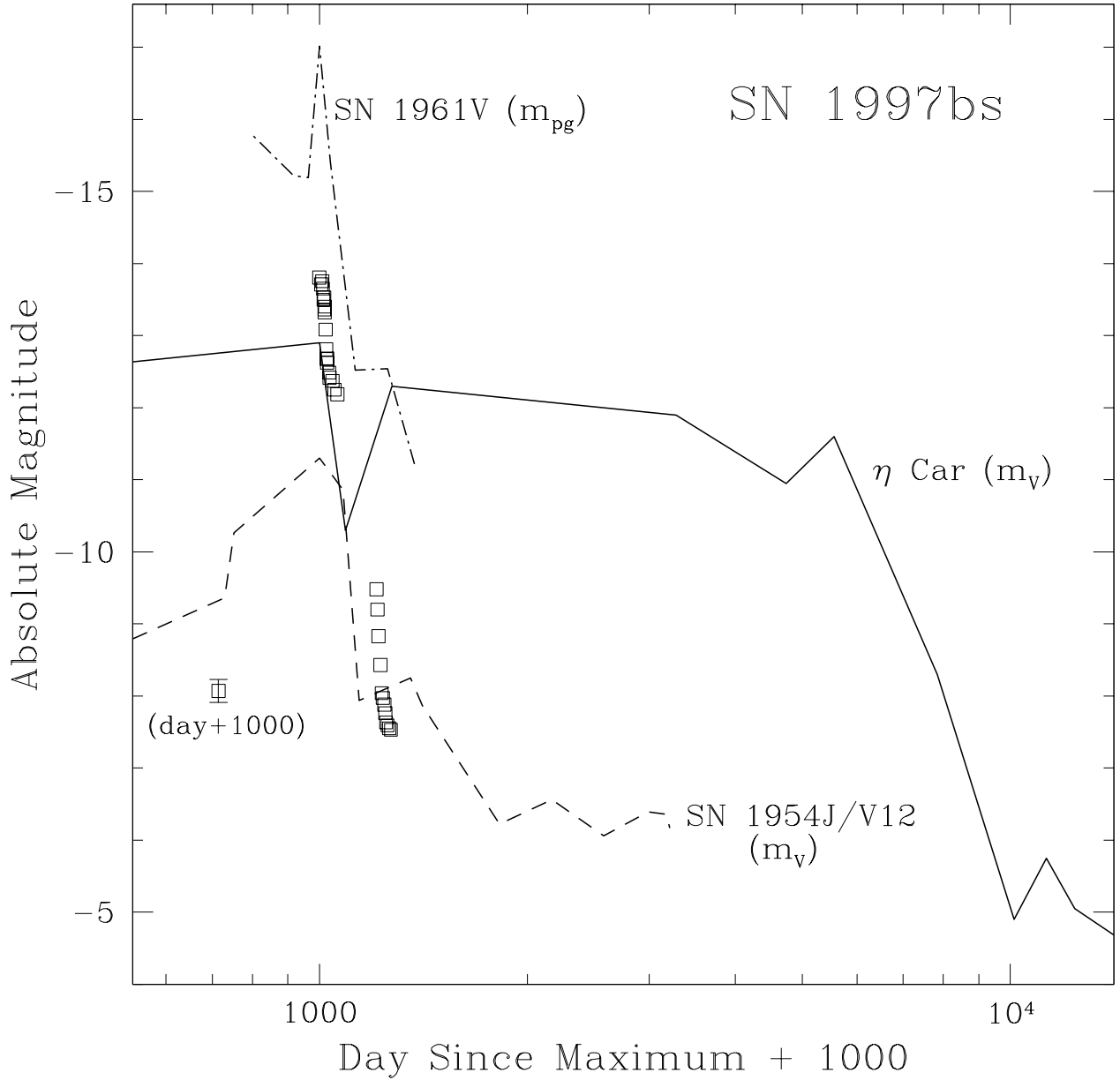}
\caption{A comparison of the light curve from Figure 5 of SN 1997bs with
those of the ``$\eta$ Car-like'' variables, SN 1961V, SN 1954J (V12 in NGC 2403),
and $\eta$ Car.  We have adapted schematic light curves for the latter three 
objects, uncorrected for extinction, from Humphreys et al.~(1999, their Figure 1) and the original sources 
for the photometry.  Absolute $V$ light curves are shown for SN 1997bs, $\eta$
Car, and SN 1954J (Tammann \& Sandage 1968);
the light curve for SN 1961V is photographic.  For SN 1997bs we also
show photometry 
for the progenitor star from an {\sl HST\/} archival WFPC2 F606W image made 
on 1994 December 28 (we have added 1000 days to this epoch in the figure).
All of the curves have been 
adjusted, so that the maxima occur at the same time.}
\end{figure}

\clearpage

\begin{figure}
\figurenum{7}
\caption{The environment of SN 1997bs (indicated by the arrow near the
center) on a coadded {\sl HST\/} archival WFPC2 F555W image.}
\end{figure}


\begin{figure}
\figurenum{8}
\plotone{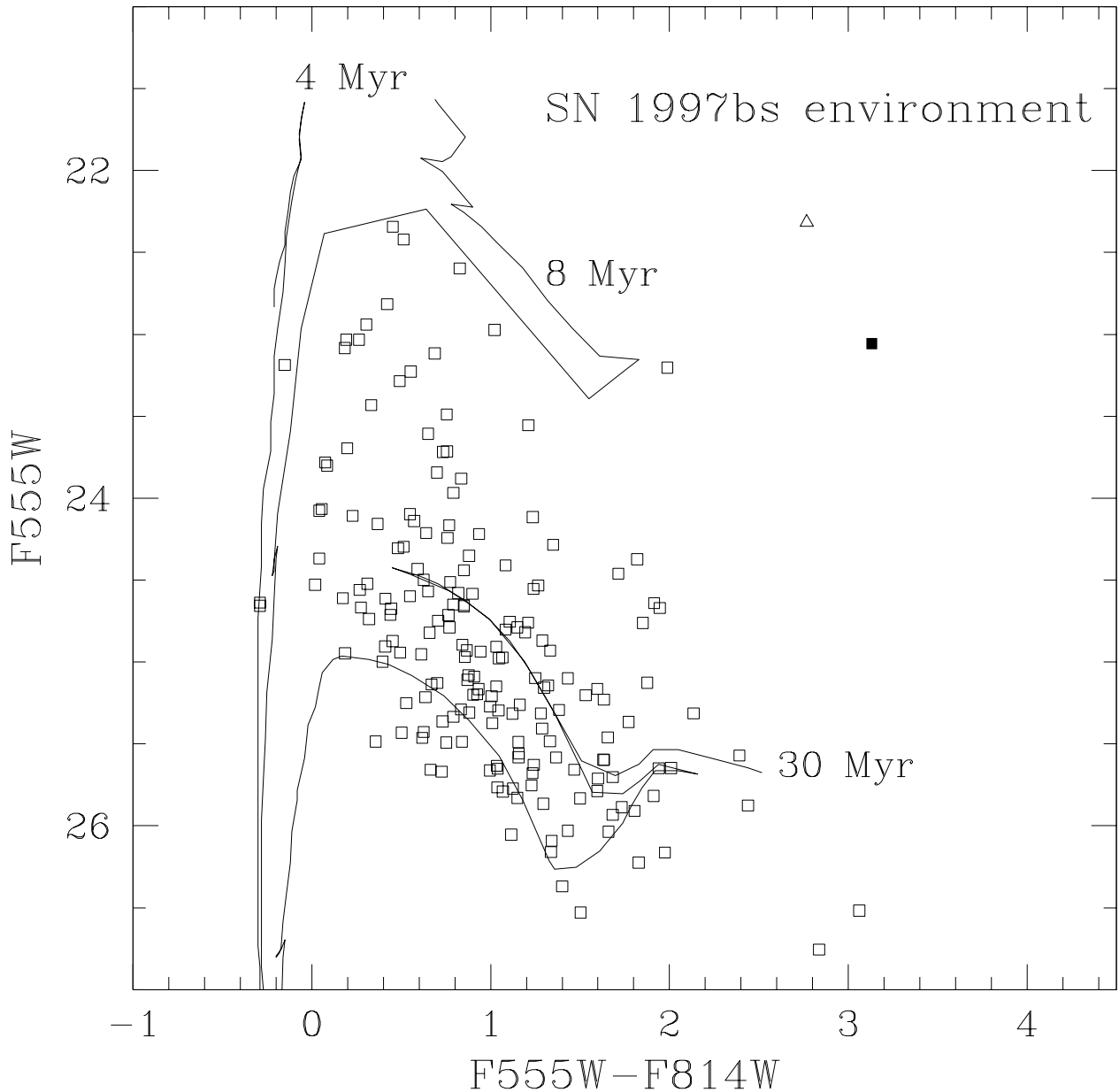}
\caption{Color-magnitude diagram of stars in the environment of SN 1997bs,
based on photometry of coadded {\sl HST\/} archival WFPC2 F555W (${\sim}V$)
and F814W (${\sim}I$)
images.  The environment represented covers 20\arcsec\ (north-south) $\times$ 
12\arcsec\ (east-west) centered on the SN.
Also shown are isochrones from Bertelli et al.~(1994) with solar
metallicity for ages 4 Myr, 8 Myr, and 30 Myr.  
The colors of the bluest stars are consistent with the Galactic foreground
reddening, $E(B-V)=0.03$ mag.
The SN is the (time-averaged) point ({\it filled square}) at 
F555W $\approx$ 23.1 and 
F555W$-$F814W $\approx$ 3 mag.  The other bright red object ({\it triangle}), 
at F555W$\approx$22.3 and
F555W$-$F814W $\approx$ 2.8 mag, is too bright to be a single star and is 
probably a red, or reddened, compact star cluster.}
\end{figure}

\end{document}